\documentstyle[12pt]{article}
\newcommand{\be}{\begin{equation}}
\newcommand{\ee}{\end{equation}}
\newcommand{\ba}{\begin{eqnarray}}
\newcommand{\ea}{\end{eqnarray}}
\newcommand{\baa}{\begin{eqnarray*}}
\newcommand{\eaa}{\end{eqnarray*}}
\newcommand{\bb}{\begin{thebibliography}} \newcommand{\eb}
{\end{thebibliography}}
\newcommand{\ci}[1]{\cite{#1}}
\newcommand{\bi}[1]{\bibitem{#1}}
\begin{document}

\begin{center}
{\Large \bf RELATIONS BETWEEN SPIN STRUCTURE FUNCTIONS
  AND QUARK MASS CORRECTIONS TO BJORKEN SUM RULE}\\[1cm]
{O.V. Teryaev}\\[0.3cm]
{\it Bogoliubov Laboratory of Theoretical Physics\\
Joint Institute for
 Nuclear Research, Dubna\\
141980 Dubna, Russia} \footnote{E--mail:
teryaev@thsun1.jinr.dubna.su}\\[1.2cm] \end{center}

\begin{abstract}

The power correction, surviving the zero quark mass limit and
found earlier to restore the validity of Burkhardt-Cottingham (BC) sum rule,
provides the contribution to the Bjorken sum rule as well.
The leading perturbative corrections to the Bjorken and Gross-Llewellyn
Smith (GLS) sum rules are increased by the factor $4/3$ and lead to
the expression
$1-4\alpha_S / 3 \pi$ instead of usual $1-\alpha_S /\pi$.
The resulting value for the Bjorken sum rule
is in excellent agreement with the most precise SLAC data, while the
observed deficit of the GLS sum rule disappear within the experimental
errors.

\end{abstract}

The structure functions $G_1$ and $G_2$ describing the spin-dependent
part of deep-inelastic scattering were discussed by Feynman in his
lectures \cite{Fey} and the definition he used is very appealing due
to the simple partonic interpretation of the dimensionless function
$g_1(x)$, the only one which survives in the scaling limit for the
longitudinal polarization case. However, when one studies the
relatively low-$Q^2$ region, the alternative definition of structure
functions, proposed by Schwinger a few years later \cite{Sch} appears
to be useful as well. In fact, it was recently applied to explain the
strong $Q^2$-dependence of the generalized Gerasimov-Drell-Hearn (GDH)
sum rule\cite{SoTe} and to clarify a number of the related problems
\cite{SoTe2}.

In the present paper this approach is applied to the first
moment of the $g_1$ structure function entering the fundamental
Bjorken sum rule. This quantity was extensively studied in
perturbative QCD and the three-loop corrections were
calculated  exactly
\ci{Lar}. As a result,
the additional perturbative {\it one-loop} correction,
originating from the non-smooth zero-quark-mass limit, is found.

The basic point of the approach \cite{SoTe,SoTe2} is the consideration
of the structure functions $g_T(x)=g_1(x)+g_2(x)$ and $g_2(x)$
as independent, while $g_1(x)$ (the most familiar one)  is expressed as:

\begin{equation}
\label{def}
g_1(x)=g_T(x)-g_2(x)
\end{equation}

In the resonance region the strong $Q^2-$dependence of $g_2$ due to the
BC sum rule naturally provides the similar dependence of the
generalized GDH sum rule \ci{SoTe}. At the same time,
the BC sum rule in the scaling region is

\begin{equation}
\label{BC}
\int^1_0 g_2(x) dx=0,
\end{equation}
and, at first sight, absolutely does not contribute to the first moment
of $g_1$:

\begin{equation}
\label{zero}
\int^1_0 g_1(x) dx=\int^1_0 g_T(x) dx.
\end{equation}

However, (\ref{BC}) happens to be valid, if and only if the specific
perturbative correction is taken into account. Consequently, this
correction should contribute to the r.h.s of (\ref{zero}).

The problem arose a couple of years ago due to the paper R. Mertig
and W.L. van Neerven{\ci{Mer-vN}, who observed that the
BC sum rule is violated in massive perturbative on-shell QCD at
one-loop level.
However, their result was in contradiction with the
QED calculation performed almost 20 years earlier by
Wu-Yang Tsai, L.L. DeRaad, Jr. and K.A. Milton \ci{Mil}\footnote{
Probably, this fact was not well known,
because the authors of \ci{Mil} used the Schwinger definitions
for the spin-dependent structure functions.}
(the result in QCD is the same apart from a trivial
color factor).
The origin of this discreapance was identified in the papers \ci{SoTe2,Mus}.
The QED calculation \ci{Mil} treated the fermion mass exactly.
The only
difference of the QCD case \ci{Mer-vN} is that the quark mass $m$ is just a
regulator of collinear singularities, and only the terms contributing to
the limit $m\rightarrow 0$ {\it before the integration over $x$}
 were taken into account. However, this leads
to the missing of the finite term at the elastic limit, when
the emitted gluon is "soft",
\begin{equation}
\delta_m^{soft} g_2 (x)=C_F{\alpha_s \over {8 \pi}} \delta_+ (1-x),
\end{equation}
restoring the validity of the BC sum rule. The same result was
obtained by G. Altarelli, B. Lampe, P. Nason and G. Ridolfi \ci{Alt}, who
derived
the exact mass-dependent formula in QCD, coinciding, up to the mentioned
colour factor, with the QED result \cite{Mil}. These authors also
established in details the origin of the extra term.

This mass correction, due to (\ref{def}) appears in the expression for
$g_1$ as well. The important point here is that the similar finite
terms for the function $g_T$ are {\it absent} \ci{Mus,Ver}.
As a result the full correction to the $g_1$ is just

\begin{equation}
\delta_m^{soft} g_1 (x) =-\delta_m^{soft} g_2 (x)=C_F{\alpha_s \over {8 \pi}}
\delta_+
(1-x).
\end{equation}

This contribution is especially important in the case of the first moment.
The partial conservation of the non-singlet axial current leads to the
zero anomalous dimension and to the manifestation of this contribution at
the leading approximation.
As a result, the correction to the Bjorken sum rule is changed:

\begin{equation}
\delta_m^{soft} \int_0^1 g_1 (x) dx=-C_F{\alpha_s \over {8 \pi}},
\end{equation}

and the full expression
for the coefficient function can be written as:

\begin{equation}
\int_0^1 g_1 (x) dx = {1 \over 2}(1-{3 \over 4}C_F{\alpha_s \over {\pi}}|_{m
\equiv 0}-{1 \over 4}C_F{\alpha_s \over {\pi}}|_{m \to 0}^{soft}).
\end{equation}

This is a principal result of this paper. However, its importance
requires further investigation.
While such a correction is due to the integration in the region of
the "soft" emitted gluons ($x \sim 1$),
there is another source of the finite mass-to-zero contribution
due to the collinear gluons. In the complete analogy to the
$\delta_+ (1-x)$ one get $\delta_+ (\theta^2)$, where $\theta$
is the gluon emission angle in the c.m. frame. As a result, there is
a finite helicity-flip cross-section in the zero-mass limit. This
effect in QED was discovered by Lee and Nauenberg in their celebrated paper
\ci{Lee} and was extensively studied and applied recently
\ci{Ein2,Seh}.

The main quantitative result is the fermion  helicity-flip probability
(analogous to the GLAP kernel, except this is finite rather than logarithmic
contribution), whose straightforward generalization
to the QCD case is:

\begin{equation}
\label{P+-}
P_{+-}(x)= C_F{\alpha_s \over {2 \pi}} (1-x).
\end{equation}



There is a similar effect in the helicity-non-flip cross section as
well. This is easily recovered by, say,  the expansion of the numerator
of (8) in \ci{Seh}:
\begin{equation}
{\theta^2 d \theta^2 \over {(\theta^2+m^2/E^2)^2}}
={1 \over {\theta^2+m^2/E^2}}-{m^2/E^2 \over {(\theta^2+m^2/E^2)^2}}
\end{equation}

Keeping $O(m^2)$ one immediately get the finite correction to the
standard logarithmic term, whose $x$-dependence is exactly the same:

\begin{equation}
\label{P++}
P_{++}(x)= - C_F{\alpha_s \over {2 \pi}} \cdot {1+x^2\over {1-x}}.
\end{equation}

The spin-averaged correction is just

\begin{equation}
\label{P}
P_a(x)= P_{++}(x)+P_{+-}(x)= - C_F{\alpha_s \over { \pi}} \cdot {x\over
{1-x}}
\end{equation}
and exactly coincide with the earlier result of Baier, Fadin
and Khoze \ci{Baier}. The agreement with this paper before the angular
integration was proved in \cite{Seh}, although the finite correction
to the helicity-conserving kernel was not presented explicitly.

Let us consider the longitudinally polarized quark. Its density matrix can
be represented as a difference of the density matrices with positive and
negative helicities. The $g_1$ structure function is proportional to:

\be
g_1(x) \sim \sigma_{++}-\sigma_{+-}-\sigma_{-+}+\sigma_{--}.
\ee

Taking into account that $\sigma_{--}=\sigma_{++}, \
\sigma_{+-}=\sigma_{-+}$, and that "collinear" contribution results in the
substitutions $\sigma_{++} \to \sigma_{++}(1+P_{++})+\sigma_{+-}
P_{+-}, \
\sigma_{+-} \to \sigma_{+-}(1+P_{++})+\sigma_{++} P_{+-}$, one should
get for the "collinear" correction to $g_1$

\begin{equation}
\label{g1c}
g_1(x)^{coll}={1\over 2}(\delta (1-x)+P_{++}(x)-P_{+-}(x))={1\over 2}
[\delta (1-x)-C_F{\alpha_s \over { \pi}}({1\over {1-x}}+x)],
\end{equation}
where the Born term is kept to make the normalization clear. Passing
to the calculation of the first moment one meet the infrared singularity
at $x=1$. It is very important, that the virtual contribution, providing
its cancelation, is proportional to the same combination
$(log(E^2/m^2)-1)$, appearing in the expression for the GLAP kernel and its
correction $P_{++}$. As a result, the $"+"$ prescription should be applicable
to $P_{++}$ in complete similarity to the logarithmic term.

\begin{equation}
\label{P+++}
P_{++}(x)= - C_F{\alpha_s \over {2 \pi}} ({1+x^2\over {1-x}})_+.
\end{equation}

The first moment of $P_{++}$ is then zero and the correction to the
Bjorken sum rule is completely determined by the helicity-flip kernel:

\begin{equation}
\int_0^1 g_1(x)^{coll} dx = {1\over 2}(1-\int_0^1 dx P_{+-}(x))={1\over 2}
(1-C_F{\alpha_s \over {4 \pi}}).
\end{equation}

Finally, the one-loop correction to Bjorken sum rule for the quark is:

\begin{equation}
\label{tot}
\int_0^1 g_1(x) dx = {1\over 2}
[1-C_F{\alpha_s \over {4 \pi}}(3(m \equiv 0)+1(soft)+1(coll))],
\end{equation}

One may worry, is it really possible just to add "zero-mass", "soft"
and "collinear" terms. The explicit calculation of $g_{1+2}$ and $g_2$
on mass shell keeping mass exactly \cite{Ver}
confirm this naive derivation.  The "collinear" contribution to $g_1$
reproduce (\ref{g1c}) while the total correction is just $-5C_F/4 \pi$,
like (\ref{tot}). The same value was obtained in \ci{Mer-vN},
(where, however, soft correction was not present) and was interpreted
as a manifestation of the regularization dependence.


Let us turn to the physical applications of these results. The status of the
soft and collinear contributions is then quite a different. The
collinear contributions correspond to the integration in the region
of the low transverse momenta ($k_T \leq m_q^2$). For the light quarks
because of the confinement the quark mass should be replaced by the
pion one \cite{GorIo}. This contribution is normally excluded by
the cuts when the experimental
data are obtained. The situation here is quite analogous to the
anomalous gluon contribution to the $g_1$ structure function
\cite{EST}. Let us note also in this connection that this naturally
explains the results obtained in \cite{Seh2}; namely, that the spin-flip
collinear effects contribute to the "normal" piece of the
spin-dependent photon spin structure function. This is due to the
fact, that these effects are related to the low transverse momenta of
scattered quarks.

For the heavy quarks, however, this correction should
be taken into account.

 From the other side, there are no reasons to exclude "soft" piece
for both heavy and light quarks. It corresponds to the integration
over x when $1-x \sim m_q^2/Q^2$. Again, the confinement effects will
change quark mass to the pion one. In principle, one can not exclude that
the accurate treatment of quark condensates instead of quark mass term
may change the actual value of the correction. Anyway, this $x$ due to the
standard convolution formula

\begin {equation}
  g_{hadron}(x_B) = \int_{x_B}^1 dx \Delta q(x) g_{quark}(x/x_B)
\end{equation}
(where $ \Delta q(x)$ is a spin-dependent quark distribution)
do not correspond to the particular value of the observed $x_B$.

It is interesting, that for the $g_2$ structure function the soft and
collinear contributions cancel each other \cite{Ver}. As a result, the
Burkhardt-Cottingham sum rule is valid either if one put quark mass
equal to zero from the very beginning, or if one take into account
all the effects, surviving the zero-mass limit (like for the heavy
quarks). If only the soft contributions is taken into account
(like for the light quarks), the BC sum rule is violated to the same
extent, up to a sign, as was reported by Mertig and van Neerven.
This sign difference is due to the fact, that in \ci{Mer-vN} the
"collinear' contribution was taken into account (it is just the
compensating terms of order $q^2/m^2$ mentioned at p.491 of this
reference), contrary to the "soft" one.

Finally, inserting the experimentally known expressions for the first
moment of the spin-dependent distributions one get the total one-loop
correction to $g_1^{p-n}$

\begin {equation}
\label{fin}
\int_0^1 g_1^{p-n}(x) dx={g_A\over 6}
(1-C_F{\alpha_s \over {\pi}})
\end{equation}

Let us discuss  briefly the experimental situation (see \cite{ALE} and
ref. therein). The multiplication
of the 1-loop correction to the Bjorken sum rule by the light-quark
factor  $4/3$ significantly improve the correspondence of the result
with the  most exact data of E143 experiment at SLAC. Namely, one get
the value $1.64 \pm 0.09$ instead of the standard one $1.72 \pm 0.09$.
If one take into account the power corrections calculated in the
framework of QCD sum rules method results in the value $1.54 \pm
0.017$ instead of $1.62 \pm 0.017$. One should compare this with the
most recent experimental value $1.51 \pm 0.013$. Although all the
theoretical values coincide with the experimental one within the
errors, the mass correction make this agreement for the central values
much better.
I especially would like to stress the perfect agreement of the
QCD sum rules calculation (note that it was obtained with only
massless perturbative corrections taken into account, when the
agreement is substantially worse).

As it is well known, the correction to the Bjorken sum rule coincide
with that for the Gross-Llewellynn Smith (GLS) sum rule
\ci{Lar,Kat}. Although
in the massive case one can not simply transform the relevant diagrams
to each other (the $\gamma^5$ moving change the signs of the mass
term \footnote{This effect spoils, in principle, the arguments of
\ci{Mer-vN}, who
used the correction to GLS sum rule calculated earlier to get the answer for
Bjorken sum rule.}),
$O(m^2)$ terms in the numerator coming from the trace are not responsible
for the extra "soft" term in $g_2$ (and, consequently, $g_1$), as it
was first mentioned in \cite{Alt}.
As a result, the coefficient function for the light quarks receives
the same contribution:
\begin {equation}
\label{GLS}
\int_0^1 F_3(x) dx = 3(1-C_F{\alpha_s \over \pi}).
\end{equation}

This relation, however, requires an additional check by the
straightforward calculation, because one can not exclude, in
principle, the appearance of such corrections coming from another
terms. This work is now in progress \cite{Ver2}.

However, the extra factor $4/3$ already completely remove
the discreapance \cite{Sidor} of the data with perturbative QCD for
all $Q^2$.
Note that such a 'compensating' factor fully respects the qualitative
nature of this
discreapance: the experimental data as a function of $Q^2$
are going above the theoretical curve and the difference is decreasing
for higher $Q^2$.

The corrections to the Bjorken and GLS sum rules are in turn related
to the correction to the $e^+ e^-$-annihilation total cross section.
This is due to the famous Crewther relation \cite{Crewter}, studied
recently for high
orders of perturbation theory\cite{Kat}. However, the zero mass limit
in this
case is known to be smooth\cite{PSF}. It seems that there is no
contradiction
here. The Crewther relation emerges due to the non-renormalization of
axial anomaly, which is not directly related to the mass contributions
studied here.

In conclusion, the new perturbative QCD correction to the partonic sum
rules is found. It comes from the quark mass contribution, having the
non-zero limit, when the quark mass tends to zero.
This correction seems to improve the agreement with the
experimental data for Bjorken sum rule. Taking into account the
similar correction removes the deficit of Gross-LLewellyn Smith sum rule.

Although the  manifestation of these correction
requires further investigation, one may conclude that the study of
mass effects (even in a zero mass limit) in hard processes
(in particular, in the partonic sum rules) seems to be very interesting.

I am thankful to A.V. Efremov, J. Ho\'rej\'si, E.A. Kuraev and
S.V. Mikhailov for stimulating discussions and valuable comments.

\bb{99}

\bi{Fey} R.P.Feynman {\it Photon-Hadron Interactions}, (Benjamin,
Reading, MA, 1972).
\bi{Sch} J.Schwinger, Proc. Natl. Acad. Sci. U.S.A. {\bf 72},
(1975) 1559.
\bi{SoTe} J.Soffer and O.Teryaev, Phys. Rev. Lett. {\bf
70}, (1993) 3373.
\bi{SoTe2} J.Soffer and O.Teryaev, Phys. Rev. {\bf D51}, (1995) 25.
\bi{Lar} S.A. Larin, J.A.M. Vermaseren, Phys. Lett {\bf B259}, (1991)
345.
\bi{Mer-vN} R. Mertig and W.L. van Neerven, Z. Phys.{\bf C60}, (1993) 489.
\bi{Mil} Wu-Yang Tsai, L.L. DeRaad, Jr. and K.A. Milton, Phys.
Rev. {\bf D11}, (1975) 3537.
\bi{Mus} I.V. Musatov, O.V. Teryaev, CEBAF Report TH-94-25(unpublished).
\bi{Alt} G. Altarelli, B. Lampe, P. Nason and G. Ridolfi, Phys. Lett
{\bf B334}, (1994) 187.
\bi{Ver} I.V. Musatov, O.V. Teryaev and O.L. Veretin, in preparation.
\bi{Lee} T.D. Lee and M. Nauenberg, Phys. Rev. {\bf B133}, (1964) 1549.
\bi{Ein2} H.F. Contopanagos and M.B. Einhorn, Nucl. Phys {\bf B377},
(1992) 20
\bi{Seh} B. Falk and L.M. Sehgal, Phys. Lett {\bf B325}, (1994) 509.
\bi{Baier} V.N. Baier, V.S. Fadin and V.A. Khoze, Nucl. Phys. {\bf
  B65}, (1973) 381.
\bi{GorIo} A.S. Gorskii, B.L. Ioffe, A.Yu. Khodzhamirian,
Phys. Lett. {\bf B227}, (1989) 474.
\bi{EST} A.V.Efremov, J.Soffer, O.V.Teryaev, Nucl.Phys. {\bf B346},
(1990) 97.
\bi{Seh2} A. Freund and L.M. Sehgal, Phys. Lett {\bf B341}, (1994) 90.
\bi{ALE} M. Anselmino, A.V. Efremov and E. Leader, CERN Preprint
CERN-TH.7216/94 (to be published in Physics Reports).
\bi{Kat} D.J. Broadhurst and A.L. Kataev, Phys. Lett {\bf B315},
(1993) 179
\bi{Sidor} A.L. Kataev and A.V. Sidorov, Phys. Lett. {\bf B331},
(1994) 179
\bi{Ver2} O.V. Teryaev, O.L. Veretin, in progress.
\bi{Crewter} R.J. Crewther, Phys. Rev. Lett. {\bf 28}, (1972) 345.
\bi{PSF}  J.Schwinger, {\it Particles, Sources and Fields}
(Addison-Wesley, 1989) v. 3, p.99.

\eb

\end{document}